\documentclass[pra, twoside, twocolumn, showpacs,eqsecnum]{revtex4}

\usepackage{amsmath,amsfonts,amssymb,amsthm}
\usepackage{graphicx}

\newcommand{\bra}[1]{\langle#1|}
\newcommand{\ket}[1]{|#1\rangle}
\newcommand{\braket}[2]{\langle#1|#2\rangle}

\newcommand{\proj}[1]{P_{#1}}
\newcommand{\ie}{i.e.~}

\newcommand{\integer}{\mathbb{Z}}
\newcommand{\real}{\mathbb{R}}

\newcommand{\hilb}{\mathcal{H}}
\newcommand{\evec}[1]{\mathbf{e}_{#1}}
\newcommand{\sumvec}[2]{\evec{#1}+\dots+\evec{#2}}

\newcommand{\addfigure}[3][width=.8\columnwidth]{
  \begin{figure}[htb]
    \begin{center}
      \includegraphics[#1]{#2}
    \end{center}
    \caption{#3}
    \label{fig:#2}
  \end{figure}
}

\begin{document}

\newtheorem{theorem}{Theorem}
\newtheorem{lemma}[theorem]{Lemma}
\newtheorem{definition}[theorem]{Definition}

\title{Quantum walks with random phase shifts}
\author{Jozef Ko\v s\'\i k$^{1,2}$,  Vladim\' \i r Bu\v{z}ek$^{1,3}$ and Mark Hillery$^{4}$}
\address{
$^{1}$Research Center for Quantum Information, Slovak Academy of Sciences,
       D\'{u}bravsk\'{a} cesta 9, 845 11 Bratislava, Slovakia\\
$^{2}${\em Quniverse}, L{\'\i}\v{s}\v{c}ie \'{u}dolie 116, 841 04 Bratislava, Slovakia\\
$^{3}$Abteilung f\"{u}r Quantenphysik, Universit\"{a}t Ulm, 89069 Ulm, Germany\\
$^{4}$Department of Physics, Hunter College of CUNY, 695 Park Avenue,  New York, NY 10021 USA
  }
\date{6 June 2006}

\begin{abstract}
  We investigate quantum walks in multiple dimensions with
  different quantum coins. We augment the model by assuming
  that at each step the  amplitudes  of the  coin state are
  multiplied by random phases. This model enables us to study in detail the role
  of decoherence in  quantum walks and to investigate the quantum-to-classical transition.
  We also provide classical
analogues of the quantum random walks studied.
   Interestingly
enough, it turns out that the classical counterparts of some
quantum random walks are classical random walks with a memory
and biased coin.
  In addition
  random phase shifts ``simplify''
  the dynamics (the cross interference terms of different paths vanish on average)
  and enable us to give a compact formula for the dispersion of such walks.

\end{abstract}

\maketitle
\section{Introduction}
The concept of quantum walks (QW) has been introduced (see Ref.~\cite{ADZ93a})
in order to explore how the intrinsically statistical character of quantum
mechanics affects statistical properties of quantum analogues of classical
random walks. In particular, an example of  a random
process is a Markov chain such that the position value $x\in X$ is iteratively
updated, given by the transition probability $P(x|y)$.

Quantum walks  have been studied in connection with novel
quantum algorithms: The instances thereof were provided in
Ref.~\cite{shenvi:052307} (the quantum walk algorithm on the
hypercube with complexity $O(\sqrt{n})$) and in Ref.~\cite{CE03a}
(the quantum  walk algorithm for  subset finding). The former 
uses the quantum walk on the hypercube, while the latter uses the
quantum walk on  bipartite graphs. Quantum walks on
bipartite graphs were analyzed in Ref.~\cite{szeg:0401053}.

Various aspects of QWs have been studied in detail recently
(for a review on QWs see Ref.~\cite{Kempe2003}). In particular,
Aharonov {\it et al.} have presented an analytic description of
discrete quantum walks on Cayley graphs~\cite{AAKV01a}.
A special case of a Cayley graph, the
line, was asymptotically analyzed in Ref.~\cite{ABN+01a}. It has been shown  that,
unlike classical random walks, the probability distribution induced by
quantum  walks is not Gaussian (with a peak around the origin of the walk),
but has two peaks at positions
$\pm\frac{n}{\sqrt2}$, where $n$ is the number of steps. As a result
the dispersion of probability distribution for quantum walks
grows quadratically, compared to linear growth for classical random walks.
The role of decoherence in quantum walks has been analyzed by Kendon {\it et al.}
Ref.~\cite{Kendon2002,Kendon2003}

Quantum walks are intrinsically deterministic processes (in the same sense
as the Schr\"{o}dinger equation is a deterministic equation). Their ``classical
randomness'' only emerges when the process in monitored (measured) in
one way or another.
Via the
measurement, one can regain a classical behavior for the
process.  For instance, by measuring the quantum coin, the
quadratic dispersion of the probability distribution reverts to
a classical, linear dispersion.
If the quantum coin is measured at
every step, then the record of the measurement outcomes singles out a
particular classical path. By averaging over all possible measurement
records, one recovers the usual classical behavior
\cite{brun:052317,BCA02b}.
Instead of measuring the quantum coin after each step,
an alternative way to regain classical randomness from a quantum walk is
to replace this coin with a new quantum coin for each flip. 

After $n$ steps of the walk one
accumulates $n$ coins that are entangled with the position of the
walking  particle. By measuring a set of $n$ quantum coins, one could
reconstruct a unique classical trajectory and by averaging
over all possible measurement outcomes,  one once again recovers the
classical result.

These two approaches  to regaining  classical behavior from the
quantum walk have been contrasted in a recent work by Brun {\it et
  al.} \cite{BCA02b}. This comparison has been studied for the particular
example of a discrete walk on the line.

In the present paper we analyze the quantum-to-classical transition
using random phase shifts on the coin register. In Sec. II we
give an introduction to the quantum walk model.
 In Sec. III (part A) we augment the
model by random phase shift dynamics and present the solution
in terms of path integrals. It turns out that  on average the
interference of amplitudes of different paths is zero and we derive
the formula for the dispersion of the mean probability distribution
in compact form. We contrast the dynamics of quantum walks with two
coins (permutation symmetric and Fourier transform) with the dynamics
of classical random walks and find an equivalence between the two
(considering the possibility that the CRW has memory and a
biased coin). In part B
of Sec. III we provide the numerical results of the problem. In particular,
we briefly analyze a situation in which phases of random shifts are distributed
according to a normal distribution
that is peaked around the phase zero and with the dispersion $\sigma$.
When the dispersion is zero,
(i.e. $\sigma=0$) we recover  the QW, while
for large $\sigma$, we obtain a uniform distribution on the interval $[-\pi,\pi
]$ and the CRW is recovered.
In between we can observe a continuous quantum-to-classical transition of quantum walks.
In Sec. IV we present our conclusions. Some technical details of the 
calculations can be found in Appendix A.

\section{QW in multiple dimensions}
Let us first define a quantum  walk in $d$ dimensions -- \ie on the lattice $\mathbb{Z}^d$.
The quantum  walk is generated by a unitary operator repeatedly
applied on a vector from a Hilbert space
$\hilb\equiv\hilb_X\otimes\hilb_D$.  The Hilbert space
$\hilb_X\equiv\text{span}\{\ket{\mathbf{x}}:\mathbf{x}\in\integer^d\}$
is called the {\it position} Hilbert space. For $\forall
\mathbf{x},\mathbf{y}\in\integer^d$
we define the usual scalar product $\mathbf{x}\cdot
\mathbf{y}\equiv\sum_{j=1}^d \mathbf{x}_j\mathbf{y}_j$ and
the norm $|\mathbf{x}|=\sqrt{\mathbf{x}\cdot \mathbf{x}}$.
In the following, the distance between the vertices is a dimensionless quantity,
with the distance between adjacent vertices equal to 1.

There are $2d$ vectors $\evec{a}\in\integer^d$ such that
$|\evec{a}|=1$.
The space
$\hilb_D\equiv\text{span}\{\ket{a}:a=1,\dots,2d\}$ is spanned by
states isomorphic to $\evec{a}$. $\hilb_D$ is called the {\it direction}
Hilbert space. In the following we set $D=\{1,\dots,2d\}$.

A single step of quantum  walk is generated by the unitary operator
$U$ such that $U=S(1\otimes C)$, where
\begin{equation}
  \label{eq:1}
    S = \sum_{\mathbf{x}\in\integer^d}\sum_{a\in
      D}\ket{\mathbf{x}+\mathbf{e}_a}\bra{\mathbf{x}}\otimes\ket{a}\bra{a}\equiv\sum_{a\in
      D} T_a\otimes P_a\; ;
\end{equation}
$P_a\equiv\ket{a}\bra{a}$ and $C$ is any unitary operator. The operator $S$
changes
the state of the position register in the direction $a$, while the coin operator
$C$ operates on the direction register. For simplicity we consider the
permutation  symmetric coin
\begin{equation}
  \label{eq:2}
  C=\left[
  \begin{matrix}
    r&t&t&\dots\\
    t&r&t&\dots\\
    \vdots&\vdots&\ddots&\dots&\\
    t&\dots&t&r
  \end{matrix}\right]\; .
\end{equation}

The quantum walk is generated by a sequence
$U^n\ket{\psi_0}$, where $\ket{\psi_0}$ is some initial state. For
simplicity, we assume
\begin{equation}
  \label{eq:3}
  \ket{\psi_0}  = \ket{\mathbf{0}}\otimes\ket{s}\; ,
\end{equation}
where $\ket{s}\equiv\frac{1}{\sqrt{|D|}}\sum_{a\in D}\ket{a}$. We also
assume the so-called Grover coin \cite{MR02a},
which is a specific instance of the permutation symmetric coin
in Eq. (\ref{eq:2}),
described by the operator
\begin{equation}
  \label{eq:6}
  C_G=2\ket{s}\bra{s}-1\; .
\end{equation}

In order to find the eigensystem of $U$, we switch to the translationally symmetric
basis \cite{AAKV01a}. We set
\begin{equation}
  \label{eq:4}
  \ket{\tilde\phi_\mathbf{k}} = \sum_{\mathbf{x}\in\integer^d}e^{i\mathbf{k}\cdot\mathbf{x}}\ket{\mathbf{x}}\; ,
\end{equation}
where $\mathbf{k}\in\real^d$. By virtue of the inverse Fourier transform we obtain
\begin{equation}
  \label{eq:5}
  \ket{\mathbf{x}} = \frac{1}{(2\pi)^d}\int_{-\pi}^\pi e^{-i\mathbf{k}\cdot\mathbf{x}}\ket{\tilde\phi_\mathbf{k}}d^d\mathbf{k}\; ,
\end{equation}
with $\ket{\tilde\phi_\mathbf{k}}$ are eigenvectors of the
translation operator in the $a$-th direction, i.e.
\begin{equation}
  \label{eq:7}
  T_a\ket{\tilde\phi_\mathbf{k}} = e^{-i\mathbf{k}\cdot\mathbf{e}_a}\ket{\tilde\phi_\mathbf{k}}\; ,
\end{equation}
where $T_a=\sum_{\mathbf{x}\in\integer^d}\ket{\mathbf{x}+\mathbf{e}_a}\bra{\mathbf{x}}$. By applying the
evolution operator, we obtain
\begin{equation}
  \label{eq:8}
  \begin{split}
    U\ket{\tilde\phi_{\mathbf{k}}}\otimes\ket{\chi}&=S\ket{\tilde\phi_{\mathbf{k}}}\otimes C\ket{\chi}\\
    &=\sum_{a\in D}e^{-i\mathbf{k}\cdot\mathbf{e}_a}\ket{\tilde\phi_{\mathbf{k}}}\otimes\ket{a}\bra{a}C\ket{\chi}\\
    &=\ket{\tilde\phi_{\mathbf{k}}}\otimes\Lambda_kC\ket{\chi}\; ,
  \end{split}
\end{equation}
where
\begin{equation}
  \label{eq:9}
  \Lambda_k = \sum_{a\in D}e^{-i\mathbf{k}\cdot\mathbf{e}_a}\ket{a}\bra{a}\; .
\end{equation}
In order to simplify the notation in what follows we will denote
projectors $\ket{a}\bra{a}$ as $P_a$.
To find the eigensystem of $U$, we need to find the eigensystem
of $\Lambda_kC$. Equivalently, we need to evaluate $(\Lambda_kC)^n\ket{\chi}$.

We first use the Grover matrix $C_G$ in  Eq.\ (\ref{eq:6}). In order to find
the power of the matrix $(\Lambda_\mathbf{k}C_G)^n$ we prove the
following lemma:
\begin{lemma}
  \label{thr:1}
  Let ${\cal D}=\{\ket{a}\}$ be the orthonormal basis of a Hilbert space and
  $C_G=2\proj{s}-1$, where $\ket{s}=\frac{1}{\sqrt{|D|}}\sum_{a\in
  D}\ket{a}$. Then
  \begin{equation*}
    \left(\proj{a}C_G\right)^n=p_n\ket{a}\bra{s}+q_n\proj{a}\; ,
  \end{equation*}
  with $p_n=\frac{2}{\sqrt{|D|}}\left(\frac{2}{|D|}-1\right)^{n-1}$ and
  $q_n=-\left(\frac{2}{|D|}-1\right)^{n-1}$.
\end{lemma}
\begin{proof}
  We denote
  $\proj{a}C_G=\frac{2}{\sqrt{|D|}}\ket{a}\bra{s}-\proj{a}=p_0\ket{a}\bra{s}+q_0\proj{a}$.
  Setting $(\proj{a}C_G)^k=p_k\ket{a}\bra{s}+q_k\proj{a}$ we get that
  $p_{k+1}=p_k\big(\frac{2}{|D|}-1\big)$ and
  $q_{k+1}=q_k\big(\frac{2}{|D|}-1\big)$. By induction we immediately
  obtain the result.
\end{proof}
From Eq.~(\ref{eq:9}) we see that with the Grover coin
\begin{equation}
  \label{eq:10}
  \begin{split}
    &(\Lambda_\mathbf{k}C_G)^n=\bigg(\sum_{a\in
    D}e^{-i\mathbf{k}\cdot\mathbf{e}_a}\proj{a}C_G\bigg)^n=\\
  &=\hspace{-1pc}\sum_{(a_1,\dots,a_n)\in
    D^n}e^{-i\mathbf{k}\cdot(\mathbf{e}_{a_1}+\dots+\mathbf{e}_{a_n})}\proj{a_1}C_G\dots\proj{a_n}C_G\; .
\end{split}
\end{equation}
By induction, the expression $\proj{a_1}C_G\dots\proj{a_n}C_G$ from
Eq.~(\ref{eq:10}) can be rewritten as
\begin{lemma}
  \label{thr:2}
  \begin{equation}
    \label{eq:11}
    \begin{split}
      &\proj{a_1}C_G\dots\proj{a_n}C_G=\frac{(-1)^n}{|D|^{(2n-1)/2}}\\
      &\times\prod_{j=1}^{n-1}(|D|\delta_{a_j,a_{j+1}}-2)(\sqrt{|D|}\ket{a_1}\bra{a_n}-2\ket{a_1}\bra{s})\; .
    \end{split}
  \end{equation}
  {\rm The product in Eq. (\ref{eq:11}) is taken to be 1 for $n=1$.}
\end{lemma}

Alternatively, in Eq. (\ref{eq:10}), the last line can be
rewritten as
\begin{equation}
  \label{eq:12}
  \begin{split}
    &\proj{a_1}C_G\dots\proj{a_n}C_G \\
    &=\hspace{-1pc}\prod_{\substack{m_1+\dots+m_k=n\\
        (a^{(1)},\dots,a^{(k)})\in D^k}}(\proj{a^{(1)}}C_G)^{m_1}\dots(\proj{a^{(k)}}C_G)^{m_k}\; ,
  \end{split}
\end{equation}
where $a^{(j)}\in D$. According to Lemma \ref{thr:1} all the terms in the
product in Eq. (\ref{eq:12}) can be expressed as the linear
combination of $\ket{a^{(j)}}\bra{s}, \proj{a^{(j)}}$. Since for
$j\neq j'\Rightarrow \braket{a^{(j)}}{a^{(j')}}=0$, we get the result
\begin{equation}
  \label{eq:13}
  \begin{split}
    &(\Lambda_\mathbf{k}C_G)^n=\sum_{\text{partition}}
      e^{-i\mathbf{k}\cdot(\mathbf{e}_{a^{(1)}}+\dots+\mathbf{e}_{a^{(k)}})} \\
    &\times\bigg(S(m_1,\dots,m_k)\ket{a_1}\bra{s}+
    T(m_1,\dots,m_k)\proj{a_n}\bigg)\; .
  \end{split}
\end{equation}
The expressions for $S$ and $T$ are given by relations
\begin{eqnarray}
  \label{eq:14}
  S(m_1,\dots,m_k)&=&p_{m_1}\dots p_{m_k}\; ;\\
  T(m_1,\dots,m_k)&=&p_{m_1}\dots p_{m_{k-1}}q_{m_k}\; ,
\end{eqnarray}
where we use the notation of Lemma \ref{thr:1}. The coefficients $m_1,\dots,m_k$
give the partitioning of the integer $n$ such that $m_1+\dots+m_k=n$.
The ``partition'' in Eq. (\ref{eq:13})  means the summation over all
such partitions.

Starting with the initial state $\ket{\psi_0}=\ket{\mathbf{0}}\otimes\ket{s}$ and using the expression
(\ref{eq:11}), we obtain
\begin{equation}
  \label{eq:15}
  \begin{split}
  &(\Lambda_\mathbf{k}C_G)^n\ket{s}=\\
  &=\sum_{(a_1,\dots,a_n)\in D^n}
  e^{-i\mathbf{k}\cdot(\mathbf{e}_{a_1}+\dots+\mathbf{e}_{a_n})}
  \frac{(-1)^{n+1}}{|D|^{(2n-1)/2}}\\
  &\times\Big[\prod_{j=1}^{n-1}(|D|\delta_{a_j,a_{j+1}}-2)\Big]\ket{a_1}\; .
  \end{split}
\end{equation}
This equation takes the singular form  for $|D|=2$ (\ie, for one-dimensional
quantum walk on line) such that only the summands for which all elements
$\{a_1,\dots,a_n\}$ are distinct, contribute to the total sum. As a
consequence, the sum in Eq. (\ref{eq:15}) when $|D|=2$ is zero for
$n>2$. But this case is special in
that the   coefficient $r$ of the Grover matrix is zero. From
now on we will consider the dimension of the lattice to be equal or larger than $2$, so
that, $|D|\geq4$.

The expression (\ref{eq:15}) is  symmetric with respect to
the permutation of elements, \ie,
$\bra{a_j}(\Lambda_\mathbf{k}C_G)^n\ket{s}$ has the same value for any
$a_j\in D, n\in\integer$.
A value of the right-hand side of  Eq.~(\ref{eq:15}) depends on the term
\begin{equation}
  \label{eq:16}
  \Xi_0(a_1,\dots,a_n)=\prod_{j=1}^{n-1}(|D|\delta_{a_j,a_{j+1}}-2)\; .
\end{equation}
Obviously, $\left|\Xi_0(a_1,\dots,a_n)\right|$ is maximal for
$a_1=\dots=a_n$. More precisely, if $a_j=a_{j+1}$ for
$j=0,\dots,n-1$ in Eq.~(\ref{eq:16}), then
\begin{equation}
  \label{eq:17}
  \left|\Xi_0(a_1,\dots,a_n)\right|=O\left((|D|-2)^k\;2^{n-k}\right)\; ,
\end{equation}
and there are $O\left(|D|^{n}\right)$ such terms in the sum of Eq.~
(\ref{eq:15}).
Now  Eq.~(\ref{eq:15}) takes the form
\begin{equation}
  \label{eq:18}
  \begin{split}
    &(\Lambda_\mathbf{k}C_G)^n\ket{s}=\\
    &=\frac{(-1)^{n+1}}{|D|^{(2n-1)/2}}\hspace{-1pc}\sum_{(a_1,\dots,a_n)\in
    D^n}e^{-i\mathbf{k}\cdot(\mathbf{e}_{a_1}+\dots+\mathbf{e}_{a_n})}\Xi_0(a_1,\dots,a_n)\ket{a_1}\; .
\end{split}
\end{equation}
In what follows we will compare the quantum walk described by
Eq. (\ref{eq:8}) with the quantum walk with random phase shifts.

\section{QW with random phase shifts}
\subsection{Analytic Results}
Quantum  walks differ from classical random walks (CRW) in many respects.
One of the main differences is that
dispersions of probability distributions of CRW grow linearly with the number of steps
while for QW the dispersions grow quadratically \cite{ABN+01a}.
In what follows we will show that  introducing random phase shifts (RPS) at each step of the evolution
causes the QW to behave more like a classical random walk.
The reduction of the QW to the CRW has been discussed
in Refs.~\cite{brun:052317,BCA02b}.
The authors of these papers have discussed
two possible routes to classical behavior for the discrete QW on a line. First, the
QW-to-CRW transition has been considered as a result of
decoherence in the quantum ``coin'' which drives the walk. Second,
higher-dimensional coins have been used to ``dilute'' the effects of quantum interference.
The position variance has been used as an indicator of classical behavior.
It has been shown
that the multi-coin walk retains the ``quantum'' quadratic
growth of the variance except in the limit of a new coin for every step,
while the walk with decoherence exhibits ``classical'' linear growth
of the variance even for weak decoherence.

In what follows we will utilize a different approach to analyze the
QW-to-CRW transition. In Ref.~\cite{BCA02b} the authors used a CP-map
on the coin degree of freedom to simulate the effects of decoherence on the
quantum walks in 1 dimension. If the CP map is pure dephasing, then
the dispersion of the probability distribution is asymptotically
linear. Our approach is new in two respects:
Firstly, we generalize the problem to an arbitrary number of dimensions;
secondly we apply a different map, a sequence of random phase shifts
on the coin.
We assume that at each step a random phase shift ($\theta_a^{(n)}$) 
described by a unitary operator
\begin{equation}
  \label{eq:27}
  R(\theta^{(n)}) = \sum_{a\in D}e^{i\theta_a^{(n)}}\proj{a}
\end{equation}
on the direction register is applied to a particle.
This procedure, in effect, is equivalent to an application
of another (random) coin on
the whole direction register.
Each random sequence
$\Sigma\equiv\{R(\theta^{(n)})\}_{n=1}^\infty$ generates a different
quantum  walk
\begin{equation}
  \label{eq:29}
  U(\theta^{(1)})\cdots U(\theta^{(n)}) = \prod_{m=1}^nS\cdot\left(1\otimes R(\theta^{(j)})C\right)\; .
\end{equation}
As we shall see later on, particular random walk 
probability distributions associated with different sequences
of  random phase shifts do not differ significantly.

What is significant on the QW-RPS is that it has mixing properties
similar to the classical random walk (the linear growth of
variance) and in particular for dimension = 2, the mean probability
distribution is exactly the same as for the classical random
walk. Before we prove both statements, we derive the formula for
the dispersion of the probability distribution for QW-RPS, using a
generalized Grover (\ie permutation symmetric) coin.

The generalized Grover  coin is (cf. Eq. (\ref{eq:2}))
\begin{equation}
  \label{eq:33}
  G_{r,t}=r\sum_{a\in D}\proj{a}+t\sum_{\substack{a,b\in D\\ a\neq b}}\ket{a}\bra{b}=(r-t)\mathbb{I}+t|D|\proj{s}\; .
\end{equation}
where $P_a=\ket{a}\bra{a}$.
This operator is unitary iff the following relations hold:
\begin{eqnarray}
  \label{eq:34}
  |r|^2+(|D|-1)|t|^2&=&1\; ; \\
  \label{eq:35}
  (|D|-2)|t|^2+r^*t+rt^*&=&0 \; .
\end{eqnarray}

The QW-RPS may be thought of as a sequence of random operators
$U(\theta)$ such that
\begin{equation}
  \label{eq:20}
  U_{r,t}(\theta) = \left[\sum_{a\in D}T_a\otimes P_a\right]\;\cdot\;(1\otimes \widehat C(\theta))\; ,
\end{equation}
and
\begin{equation}
  \label{eq:21}
  \widehat C(\theta) = 1\otimes\sum_{a\in D}e^{i\theta_a}\proj{a}G_{r,t}\; .
\end{equation}
Here $\theta=(\theta_1,\dots,\theta_{|D|})$ is a sequence of
independent random real variables. Hence we actually get a
sequence of random operators $\{U(\theta^{(j)})\}_j$ which creates the
QW-RPS.

One step of a QW-RPS is given by:
\begin{equation}
  \label{eq:22}
  \begin{split}
    U_{r,t}(\theta^{(1)})&=S\cdot\widehat C(\theta^{(1)})=\Big(\sum_{a_1\in
      D}T_{a_1}\otimes\ket{a_1}\bra{a_1}\Big)\\
    &\times\Big[1\otimes\sum_{a\in
      D}e^{i\theta^{(1)}_{a}}({t}{\sqrt{|D|}}\ket{a}\bra{s}+(r-t)P_a)\Big]\\
    &=\sum_{a_1\in D}T_{a_1}\otimes
      e^{i\theta^{(1)}_{a_1}}\Big({t}{\sqrt{|D|}}\ket{a_1}\bra{s}+(r-t)P_{a_1}\Big)\; .
  \end{split}
\end{equation}
For the chain of $n$ evolution operators of QW-RPS we obtain
\begin{lemma}
  Let $\{U_{r,t}(\theta^{(j)})\}_{j=1}^\infty$ be a sequence of random
  operators according to Eq. (\ref{eq:20}). Then
  \label{thr:4}
  \begin{eqnarray}
    \label{eq:36}
    \begin{split}
      &U_{r,t}(\theta^{(1)})\cdots U_{r,t}(\theta^{(n)})=
      \sum_{a_1,\dots,a_n\in
      D}T_{a_1+\dots+a_n}\\
    &\otimes e^{i(\theta^{(1)}_{a_1}+\dots+\theta^{(n)}_{a_n})}\Big[(r-t)\ket{a_1}\bra{a_n}+{t}{\sqrt{|D|}}\ket{a_1}\bra{s}\Big] \\
    &\times\Xi(a_1,\dots,a_n)\ket{a_1}\; ,
    \end{split}
  \end{eqnarray}
  where
  \begin{equation}
    \label{eq:38}
    \Xi(a_1,\dots,a_n)\equiv\prod_{j=1}^{n-1}\Big[{t}+(r-t)\delta_{a_j,a_{j+1}}\Big]\; .
  \end{equation}
\end{lemma}
\begin{proof}
  By induction.
\end{proof}

It can be analyzed how a specific QW-RPS evolves given a
 specific initial state
$\ket{\psi_0}=\ket{\mathbf{0}}\otimes\ket{s}$.
\begin{lemma}
  \label{thr:3}
  Let $\{U(\theta^{(j)})\}_{j=1}^\infty$ be a sequence of random
  operators according to Eq. (\ref{eq:20}). Then
  \begin{equation}
    \label{eq:23}
    \begin{split}
      &\ket{\psi(\Theta,n)}\equiv U(\theta^{(1)})\cdots
      U(\theta^{(n)})\ket{\mathbf{0}}\otimes\ket{s}=\\
      &=\sum_{a_1,\dots,a_n\in
        D} \ket{\mathbf{0}+\mathbf{e}_{a_1}+\dots+\mathbf{e}_{a_n}}\otimes \\
      &\otimes e^{i(\theta^{(1)}_{a_1}+\dots+\theta^{(n)}_{a_n})}\Xi(a_1,\dots,a_n)\Big[\frac{r-t}{\sqrt{|D|}}+\sqrt{|D|}t\Big]\ket{a_1} \; .
    \end{split}
  \end{equation}
\end{lemma}
\begin{proof}
  From Lemma \ref{thr:4}.

\end{proof}
The probability distribution of $\ket{\psi(\Theta,\mathbf{x},n)}$ shall be derived by projecting it onto
$P_\mathbf{x}\otimes1$ and tracing over the coin Hilbert
space. Hence by setting
\begin{eqnarray}
\mathbf{a}&\equiv&(a_1,\dots,a_n)\\
\mathbf{a}'&\equiv&(a_1',\dots,a_n')\\
\theta(\mathbf{a})&\equiv&(\theta_{a_1}+\dots+\theta_{a_n})
\end{eqnarray}
we get
\begin{equation}
  \label{eq:24}
  \begin{split}
    P(\Theta,&\mathbf{x},n)=\|(P_\mathbf{x}\otimes1)\psi(\Theta,n)\|^2=
    \Big|\frac{r-t}{\sqrt{|D|}}+\sqrt{|D|}t\Big|^2\times\\
    &\times\sum_{\substack{\mathbf{a},\mathbf{a}'\in
      D^n\\\mathbf{a}\equiv\mathbf{x}\equiv\mathbf{a}'}}e^{i(\theta(\mathbf{a})-\theta(\mathbf{a}'))}\Xi(\mathbf{a})\Xi(\mathbf{a}')^*\braket{a_1}{a_1'}\; ,
    \end{split}
\end{equation}
where $\theta(\mathbf{a})$ is the sum of the sequence of  independent random
variables for each $\mathbf{a}$ and $\Xi$ is the same as in Eq. (\ref{eq:16}).
The symbol $\mathbf{a}\equiv\mathbf{x}$ means that $\mathbf{0}+\sum_{j=1}^n\mathbf{e}_{a_j}=\mathbf{x}$.
It is clear that $P(\Theta,\mathbf{x},n) = P(\Theta,\mathbf{x},n)^*$
since we are summing over all tuples $\mathbf{a},\mathbf{a}'$.

Eq. (\ref{eq:24}) depends on $\Theta$, which is an event
generating $n|D|$ random variables
$\{\theta_{a_j}: j=1,\dots,n,\;a\in\mathbf{a}\}$. We
can split Eq. (\ref{eq:24}) into two parts:
\begin{equation}
  \label{eq:26}
  \begin{split}
    P(&\Theta,\mathbf{x},n)=\Big|\frac{r-t}{\sqrt{|D|}}+\sqrt{|D|}t\Big|^2\Big\{\sum_{\substack{\mathbf{a}\in  D^n\\\mathbf{a}\equiv\mathbf{x}\equiv\mathbf{a}'}}|\Xi(\mathbf{a})|^2+ \\
  &+\sum_{\substack{\mathbf{a}\neq\mathbf{a}'\\\mathbf{a}\equiv\mathbf{x}\equiv\mathbf{a}'}}e^{i(\theta(\mathbf{a})-\theta(\mathbf{a}'))}
  \Xi(\mathbf{a})\Xi(\mathbf{a}')^*\braket{a_1}{a_1'}\Big\}
\end{split}
\end{equation}
To obtain the {\it mean} probability distribution, we integrate over the random
variable $\theta$,assuming uniform distribution for all phases of $\theta$:
\begin{equation}
  \label{eq:25}
  \begin{split}
    \langle P(\Theta,\mathbf{x},n)\rangle_\Theta &=
    \int_0^{2\pi}\frac{d^{|D|n}\theta}{(2\pi)^{|D|n}}P(\Theta,\mathbf{x},n)\\
    &=\Big|\frac{r-t}{\sqrt{|D|}}+\sqrt{|D|}t\Big|^2\sum_{\substack{\mathbf{a}\in D^n\\\mathbf{a}\equiv\mathbf{x}\equiv\mathbf{a}'}}|\Xi(\mathbf{a})|^2\; .
  \end{split}
\end{equation}
The terms coming from
$\sum_{\substack{\mathbf{a}\neq\mathbf{a}'\\\mathbf{a}\equiv\mathbf{x}\equiv\mathbf{a}'}}$
cancel out.
Now the mean probability distribution depends only on the term
$\Xi(\mathbf{a})$. For $d=2$ and Grover coin (\ie
$r=\frac{2}{|D|}-1,\;t=\frac{2}{|D|}$), this term is the product of
$\pm2$, and $|\Xi(\mathbf{a})|^2=2^{-2(n-1)}$. Then
\begin{equation}
  \label{eq:31}
  \langle P(\Theta,\mathbf{x},n)\rangle_\Theta =
  \frac{1}{4^n}\sum_{\substack{\mathbf{a}\in
  D^n\\\mathbf{a}\equiv\mathbf{x}}} 1\; .
\end{equation}
The sum over all paths in Eq. (\ref{eq:31}) is the same as the sum of
all classical paths. The constant is the product of the probability to
take any individual direction at each step. Hence the mean probability
distribution of QW-RPS for dimension 2 is the same as the CRW in 2
dmensions.
We easily show that the probability distribution resulting from Eq. (\ref{eq:31}) is normalized:
\begin{equation}
  \label{eq:32}
  \sum_{\mathbf{x}\in\integer^d}\langle
  P(\Theta,\mathbf{x},n)\rangle_\Theta=\frac{1}{4^n}\sum_{\mathbf{a}\in D^n}1=1 \; .
\end{equation}
It is  an interesting question whether the equivalence of QW-RPS with
CRW in two dimensions is merely a coincidence, or whether the same result applies
for higher dimensions with a modified Grover coin.

Now we arrive
at the conclusion that the averaged probability distribution of QW-RPS with
the generalized Grover coin $G_{r,t}$, is
\begin{equation}
  \label{eq:39}
  \langle
  P(\Theta,\mathbf{x},n)\rangle_{\Theta}=\Big|\frac{r-t}{\sqrt{|D|}}+\sqrt{|D|}t\Big|^2\,\sum_{\substack{\mathbf{a}\in
  D^n\\\mathbf{a}\equiv\mathbf{x}}}
    |\Xi(\mathbf{a})|^2\; .
\end{equation}
This equation is one of the main
results of our paper, as it shows that the probability
distribution of a QW-RPS corresponds to a classical sum over
paths.

In order to study specific properties of the mean probability distribution
let us consider its dispersion that is defined as
\begin{equation}
  \label{eq:28}
  \mathcal{D}(D,n) = \sum_{\mathbf{x}\in\integer^d} \langle P(\Theta,\mathbf{x},n)\rangle_\Theta|\mathbf{x}|^2.
\end{equation}

The dispersion corresponding to the mean probability  distribution is
evaluated in  Appendix A [see Eq.~(\ref{eq:62})] from which
we can derive the following theorem:
\begin{theorem}
  \label{thr:7}
  The dispersion of QW-RPS with a generalized Grover coin with
  coefficients $r$,$t$ for $n>2$ is
  \begin{equation}
    \label{eq:63}
    \mathcal{D}(D,n)=\frac{1+|r|^2-|t|^2}{1-|r|^2+|t|^2}(n-2)+O\Big((|r|^2-|t|^2)^n\Big)
  \end{equation}
\end{theorem}
We want to find the coefficients $r,t$ such that the above equation
is identical to the probability distribution of a CRW. The sufficient
condition is that all the terms in $\Xi$ have the same absolute
values. It is obvious that this is equivalent with the following requirement
\begin{theorem}
  \label{thr:5}
  The dispersion of QW-RPS with Grover coin $G_{r,t}$ is identical
  to that of a CRW if and only if $|r|=|t|$.
\end{theorem}
Comparing this requirement with Eqs. (\ref{eq:34}) -- (\ref{eq:35})  it follows that
the condition $|r|=|t|$ is satisfied only in two cases:
\begin{itemize}
\item dimension = 1, $r=\frac{1}{\sqrt2}e^{i\alpha},\;t=\frac{1}{\sqrt2}e^{i\beta},\;\alpha-\beta=\frac{\pi}{2}+k\pi$
\item dimension = 2, $r=\frac{1}{2}e^{i\alpha},\;t=\frac{1}{2}e^{i\beta},\;\alpha-\beta=\pi+2k\pi$
\end{itemize}

The average probability distribution of QW-RPS with generalized Grover
coefficients $r,t$ is actually identical to the  probability
distribution of a CRW with memory (CRW-M). We define a CRW-M as the random
walk of a particle on a $d$-dimensional lattice, whose direction is
changed at each step, depending on the direction from which it
came. The CRW-M is given by the sequence $\{(\mathbf{x}_n,
a_n)\}_{n=1}^\infty$, where $\mathbf{x}_n\in\integer^d$ is the
position of the particle,
$\mathbf{e}_{a_n}\equiv\mathbf{x}_n-\mathbf{x}_{n-1}$ is the unit
vector in any of the $2d$ directions, and $a_0$ is preset. One
step of the CRW-M is given by the transformation
$(\mathbf{x}_n,a_n)\rightarrow(\mathbf{x}_{n+1},a_{n+1})$ such that
$\mathbf{x}_{n+1}-\mathbf{x}_n=\mathbf{e}_{a_{n+1}}$, where
\begin{equation}
  \text{Prob}(a_{n+1}|a_n)=
  \begin{cases}
    \;|r|^2:& a_{n+1}=a_n\\
    \;|t|^2:& \text{otherwise}
  \end{cases}
\end{equation}
Beginning with $\mathbf{x}_0=\mathbf{0}$ and $a_0$ in the uniform
mixture of $|D|=2d$ directions, the probability
$\text{Prob}(\mathbf{x}_n=\mathbf{y})$ is given by the sum of all
paths from $\mathbf{0}$ to $\mathbf{y}$, each weighed by the product
of terms $|r|^2, |t|^2$, depending on whether the path continues in
the same direction for two consecutive steps. The sum of the
amplitudes is the same as in  Eq. (\ref{eq:39}), weighted by
the factor $\frac{|r-t|^2}{|D|}=\frac{1}{|D|}$, which corresponds to
the mixture of different values of the initial direction of the walker $a_0$
(see Eqs. (\ref{eq:34}), (\ref{eq:35})).

The memory effect in the QW-RPS with Grover coin is due to different absolute
values of the coefficients $r,t$. Although Eq. (\ref{eq:39}) shows that a QW-RPS using a
symmetric coin yields a probability distribution which corresponds to
a CRW with memory (QRW-M), it is also interesting to consider
what QW-RPSs correspond to CRWs with no memory.
Using the Fourier coin instead, and using
random
phase shifts will give the mean probability distribution equivalent to a CRW.

The Fourier coin is defined by the operator of a $d$-dimensional Fourier
transform
\begin{equation}
    F_d \equiv\frac{1}{\sqrt{d}}\sum_{j=0}^{d-1}\sum_{k=0}^{d-1}e^{2\pi ijk/d}\ket{j}\bra{k}
\end{equation}
One step of $d$-dimensional QW-RPS with Fourier coin is defined by unitary operator
\begin{equation}
    \label{eq:64}
  U_F(\theta) = \left[\sum_{a\in D}T_a\otimes P_a\right]\;\cdot\;(1\otimes \widehat F(\theta))\; ,
\end{equation}

with
\begin{equation}
\widehat{F}(\theta)=1\otimes\sum_{a\in D}e^{i\theta_a}P_aF_{|D|} \; .
\end{equation}
Obviously,
\begin{equation}
    \label{eq:65}
    \begin{split}
        U(\theta^{(1)})\dots U(\theta^{(n)})&=\sum_{a_1,\dots,a_n\in D}T_{a_1+\dots+a_n}
        e^{i(\theta^{(1)}_{a_1}+\dots+\theta^{(n)}_{a_n})}\\
        &\otimes\big(P_{a_1}F_{|D|}\dots P_{a_n}F_{|D|}\big)
    \end{split}
\end{equation}
As before, we study the mean probability distribution induced by QW-RPS with Fourier coin and we conclude that
the cross terms to vanish. It is straightforward to prove (cf.~Lemma \ref{thr:2}):
\begin{lemma}
\label{thr:8}
\begin{equation}
\begin{split}
P_{a_1}F_{|D|}&\dots P_{a_n}F_{|D|}\ket{s}=\frac{\delta_{a_n,|D|}}{|D|^{(n-1)/2}}\\
    &\times e^{2\pi i(a_1a_2+\cdots+a_{n-1}a_{n})/{|D|}}\ket{a_1}
    \end{split}
    \end{equation}
\end{lemma}
Now the probability distribution after $n$ steps, with the initial state
$\ket{\mathbf{0}}\otimes\ket{s}$ and the sequence $\Theta$ of random phases
is (cf. Eq. \ref{eq:24})
\begin{equation}
    \label{eq:66}
    \begin{split}
    P_F(\Theta,\mathbf{x},n)&=\sum_{\substack{\mathbf{a}\in D^n\\ \mathbf{a}\equiv\mathbf{x}}}
    |\Xi_F(\mathbf{a})|^2\\
    &+\sum_{\substack{\mathbf{a}\neq\mathbf{a}'\\ \mathbf{a}\equiv\mathbf{x}\equiv\mathbf{a}'}}
    e^{i(\theta(\mathbf{a})-\theta{\mathbf{a}'})}\Xi_F(\mathbf{a})\Xi_F(\mathbf{a}')^*\braket{a_1}{a_1'} \; ,
    \end{split}
\end{equation}
where
\begin{equation}
    \label{eq:67}
    \Xi_F(a_1,\dots,a_n)=\frac{\delta_{a_n,|D|}}{|D|^{(n-1)/2}}
    e^{2\pi i(a_1a_2+\cdots+a_{n-1}a_{n})/{|D|}}\; .
\end{equation}
By averaging over all sequences of random phases $\Theta$ (cf. Eq.~(\ref{eq:25}))
the second term in Eq. (\ref{eq:66}) vanishes and we get
\begin{equation}
    \label{eq:68}
    \langle P_F(\Theta,\mathbf{x},n)\rangle_\Theta=
    \frac{1}{|D|^{n-1}}\sum_{\mathbf{a}\equiv\mathbf{x}}\delta_{a_n,|D|} \; .
\end{equation}
Notice that despite the symmetric initial coin state $\ket{s}$, there is an asymmetry
manifested in the term $\delta_{a_n,|D|}$. This is due to the fact that after the
initial coin toss, the coin register is in the state $F_{|D|}\ket{s}=\ket{|D|}$.
We can symmetrize the evolution by taking the initial coin state to be
$F_{|D|}^\dagger\ket{s}=\ket{|D|}$. Then we can compute that the  probability
distrubution has the same form as Eq. (\ref{eq:66}), with $\Xi_F$ replaced by
\begin{equation}
    \label{eq:69}
    \Xi_{F,\text{sym}} = \frac{1}{|D|^{n/2}}e^{2\pi i(a_1a_2+\dots+a_{n-1}a_n)}\ket{a_1}
\end{equation}
yielding the mean probability distrubution
\begin{equation}
    \label{eq:70}
    \langle P_{F,\text{sym}}(\Theta,\mathbf{x},n)\rangle_\Theta=
    \frac{1}{|D|^{n}}\sum_{\mathbf{a}\equiv\mathbf{x}}1 \; .
\end{equation}
This is exactly the form for the probability distribution of a (memoryless) CRW,
the sum being over all paths from $\mathbf{0}$ to $\mathbf{x}$, weighed by
$\frac{1}{|D|}$.

\subsection{Numerical Results}
To complement our analytical results, we plot the dispersion of the
average probability distribution of the QW-RPS for the Grover coin:
\begin{equation}
  \label{eq:30}
  \mathcal{D}_0(D,n) =
  \Big\langle\sum_{\mathbf{x}\in\integer^d}|\mathbf{x}|^2\mathrm{Tr}\big(\proj{\mathbf{x}}
  \ket{\psi(\Theta,n)}\bra{\psi(\Theta,n)}\big)\Big\rangle_\Theta \; .
\end{equation}

The  results are shown in the following figures:
in Fig.~1 we plot the dispersion (\ref{eq:30}) as a function of the
number of steps for a two-dimensional ($d=2$)
QW, CRW and QW-RPS, respectively.

\addfigure[angle=-90,width=\columnwidth]{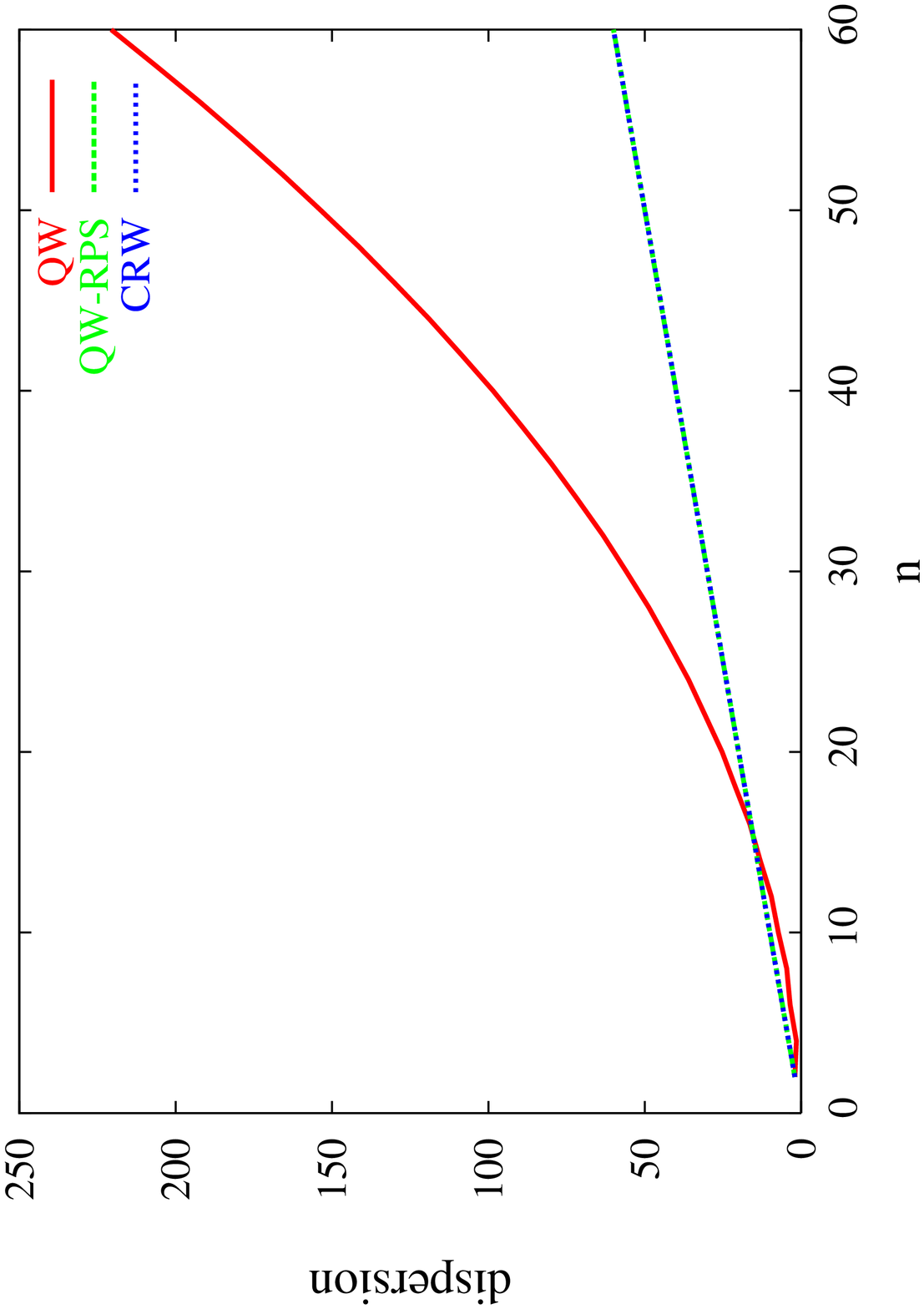}
{(Color online) Dispersions of the probability distributions corresponding
  to the $n$ steps of quantum walk (solid/red line),
  the (memoryless) classical random walk with equal probabilities of step in
  any direction (dotted/blue
  line), and the quantum walk with random phase shifts (dashed/green line)
  for a two-dimensional system.  The initial state of the quantum system system
  is described by a vector $\ket{\psi_0}=\ket{0}\otimes\ket{s}$ and we
  assume the Grover coin $C_G$. The quantities for QW-RPS were
  obtained by generating 50 evolutions of QW-RPS with respective
  dispersions of probability distributions, and by averaging over them.}

Comparing the three corresponding lines we find
that the dispersion in the QW grows quadratically with number of steps
\cite{meyer1996} (see solid/red line in Fig.~1).
This is in a sharp
contrast to a classical random walk for which the dispersion is a
linear function of the number of steps
(see dotted/blue line in Fig.~1). The lines for QW-RPS and CRW
overlap. The dispersion of either grows linearly with the number of
steps $n$ (see dashed/blue line in Fig.~1).

In Fig.~2 we plot the dispersion (\ref{eq:28}) as a function of
number of steps for a three-dimensional ($d=3$) QW, CRW and
QW-RPS, respectively.
\addfigure[angle=-90,width=\columnwidth]{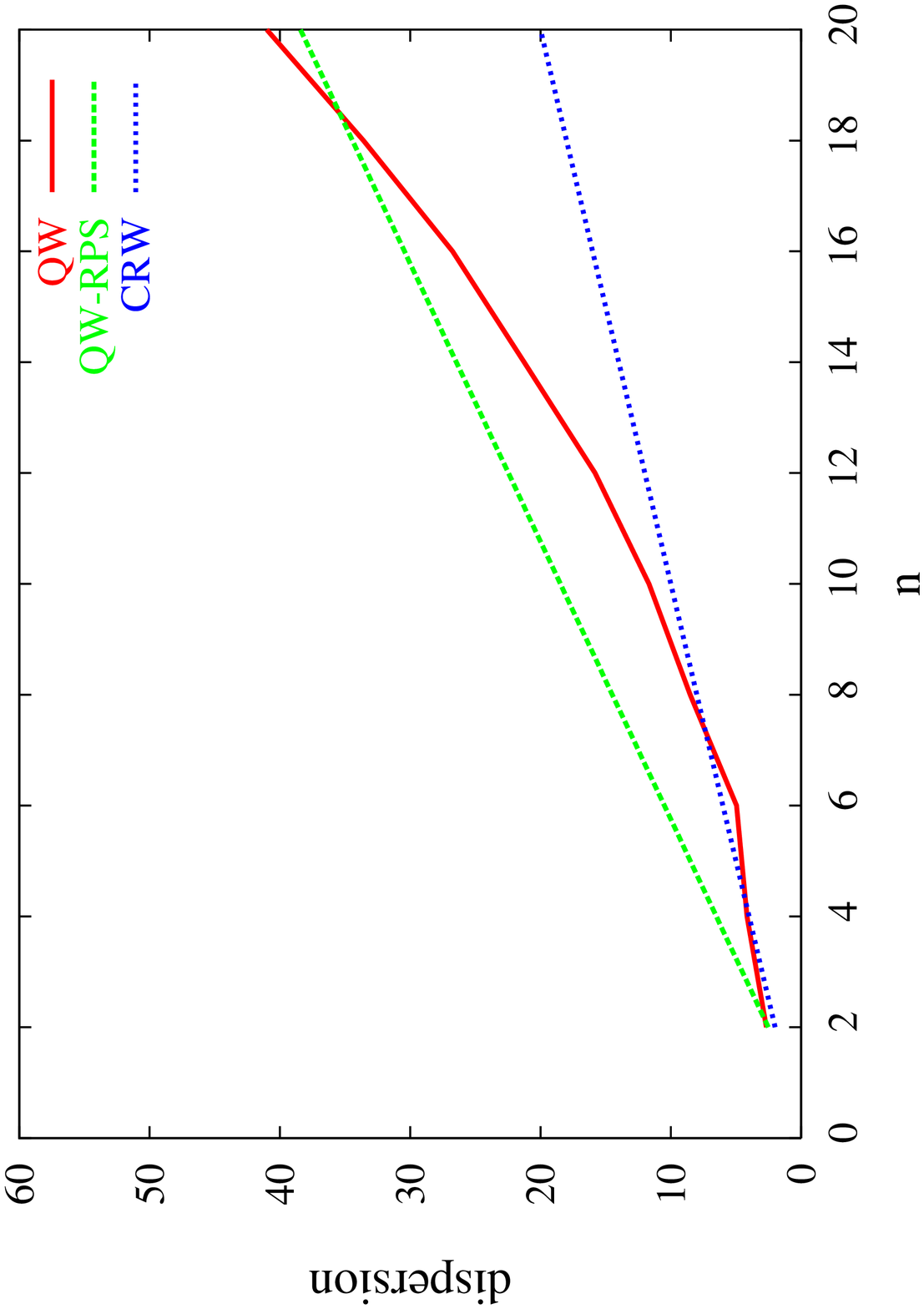} {(Color
online)
  Dispersions of probability distributions corresponding to the $n$
  steps of quantum walk (solid/red line),
  the (memoryless) classical random walk with equal probabilities of step in
  any direction (dotted/blue
  line), and the quantum walk with random phase shifts (dashed/green line)
  for a three-dimensional system.  The initial state of the quantum system system
  is described by a vector $\ket{\psi_0}=\ket{0}\otimes\ket{s}$ and we
  assume the Grover coin $C_G$.
  The quantities for QW-RPS were
  obtained by generating 50 evolutions of QW-RPS with respective
  dispersions of probability distributions, and by averaging over them.}

As in the two-dimensional case, the dispersion of the probability
distribution of the QW grows quadratically.
The dispersion of the classical random walk is a linear function of
number of steps and it does not depend
on the dimension of the random walk. Interestingly enough, the
dispersion of the quantum walk with random phase shifts
is again a linear function, but unlike in the two-dimensional case,
for $d\geq 3$ the linear growth of the dispersion is
faster than in the classical case.

The same conclusions can be derived from our simulations of
quantum walks in four-dimensional space (see Fig.~3).
\addfigure[angle=-90,width=\columnwidth]{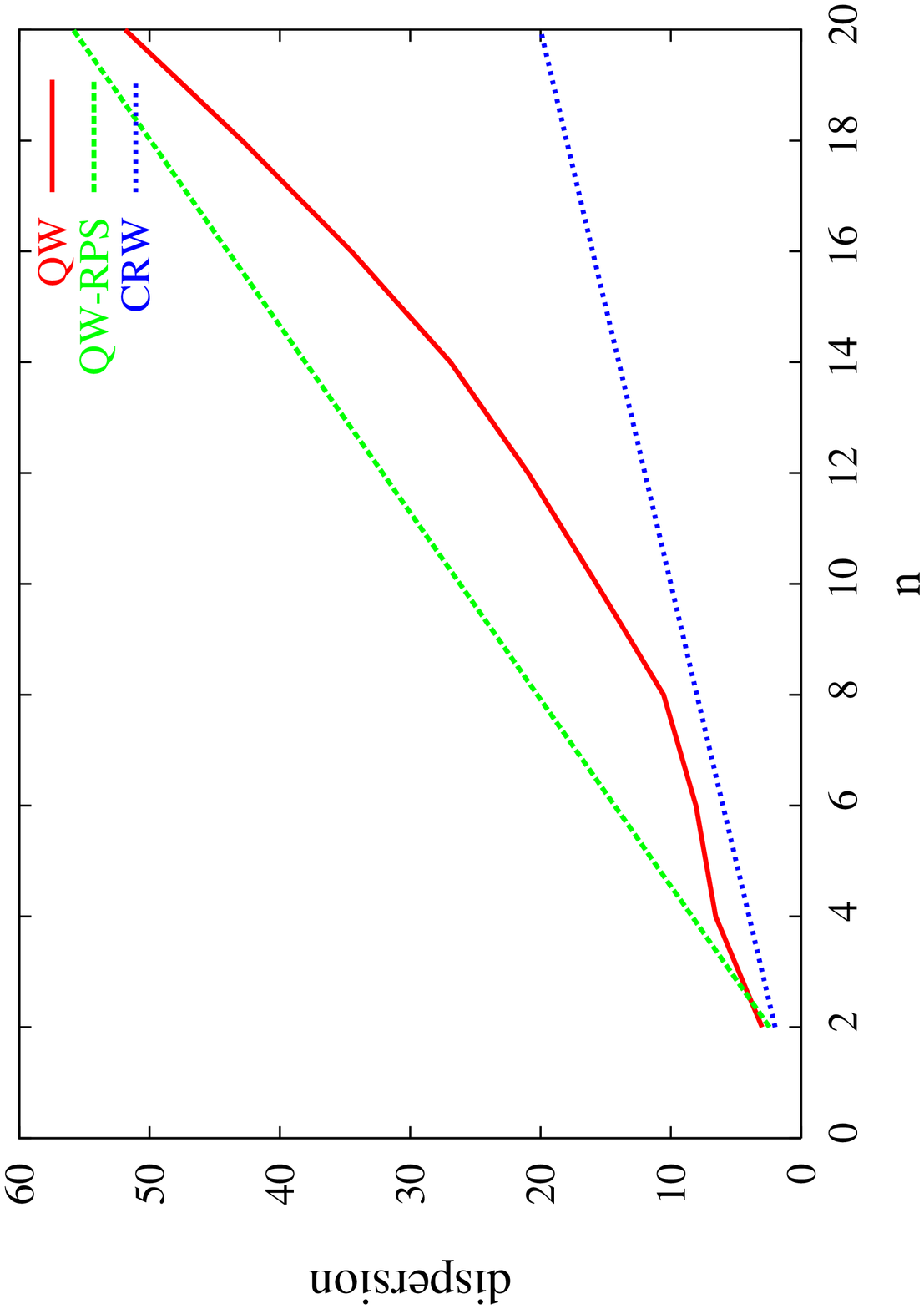} {(Color
online) Dispersions of probability distributions corresponding to
$n$ steps of quantum walk (solid/red line),
 the (memoryless) classical random walk with equal probability of step in any direction
 (dotted/blue line), and the
 quantum walk with random phase shifts ($\mathcal{D}_0(D,n)$: dashed/green line)
 for a four-dimensional system.  The initial state of the quantum system
is described by a vector $\ket{\psi_0}=\ket{0}\otimes\ket{s}$. We
assume the Grover coin $C_G$.The quantities for QW-RPS were
  obtained by generating 50 evolutions of QW-RPS with respective
  dispersions of probability distributions, and by averaging over them.}

In Fig.~4 we plot dispersion of probability distributions for quantum walks with random phase shifts
as a number of steps for various dimensions $d=2,3,4$. We generated 50
evolutions of QW-RPS for each dimension, and averaged over the
respective dispersion generated by each evolution.

We can conclude, that as the
dimension increases, the linear growth of the dispersion  also
increases.
\addfigure[angle=-90,width=\columnwidth]{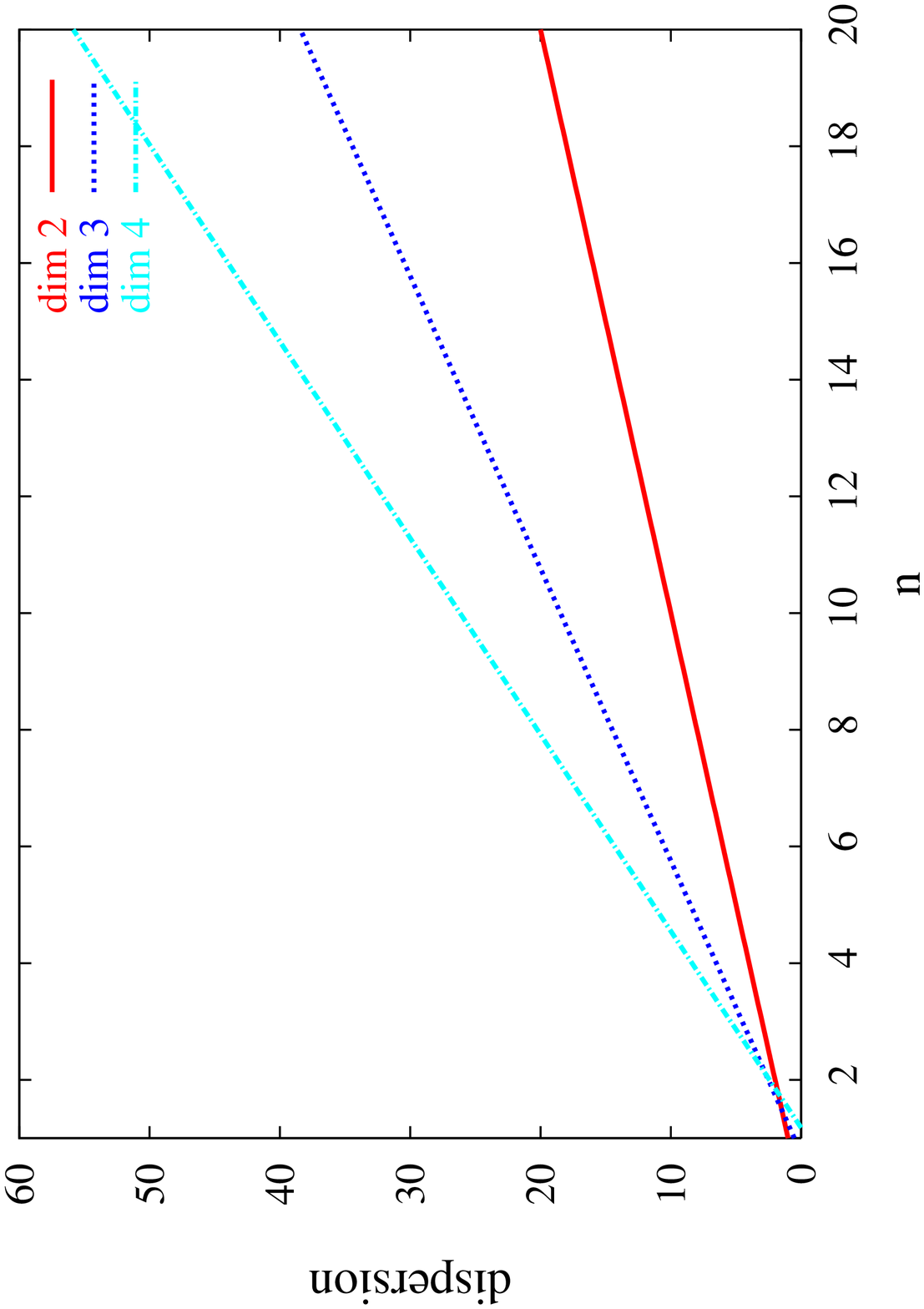} {(Color
online) Dispersion of QW-RPS processes for different dimensions
for $n$ steps. We see that these dispersions $\mathcal{D}_0(D,n)$
are linear functions with gradients that depend on the
dimensionality of the system under consideration. Only the case
$d=2$ coincides with the classical random walk. The quantities for
QW-RPS were obtained by generating 50 evolutions of QW-RPS with
respective dispersions of probability distributions, and by
averaging over them.}

We have shown that the introduction of random phase shifts causes
the transition of a QW to a (quasi-)classical random walk. In our
previous discussion we have considered random phases to be
uniformly distributed in the interval $[-\pi,\pi ]$. Here we
briefly analyze a situation when phases of random shifts are
distributed according to a normal distribution that is peaked
around the phase zero and with the dispersion $\sigma$. When the
dispersion is zero, (i.e. $\sigma=0$) we recover  the QW (see
Fig.~5), while for large $\sigma$, we obtain uniform distribution
on the interval $[-\pi,\pi ]$ and the CRW is obtained. The results
are shown in Fig.~5. This analysis clearly shows the
quantum-to-classical transition for quantum walks which is
generated by random phase shifts. As the phase shifts become more random, the walk becomes more
classical.
\addfigure[angle=-90,width=\columnwidth]{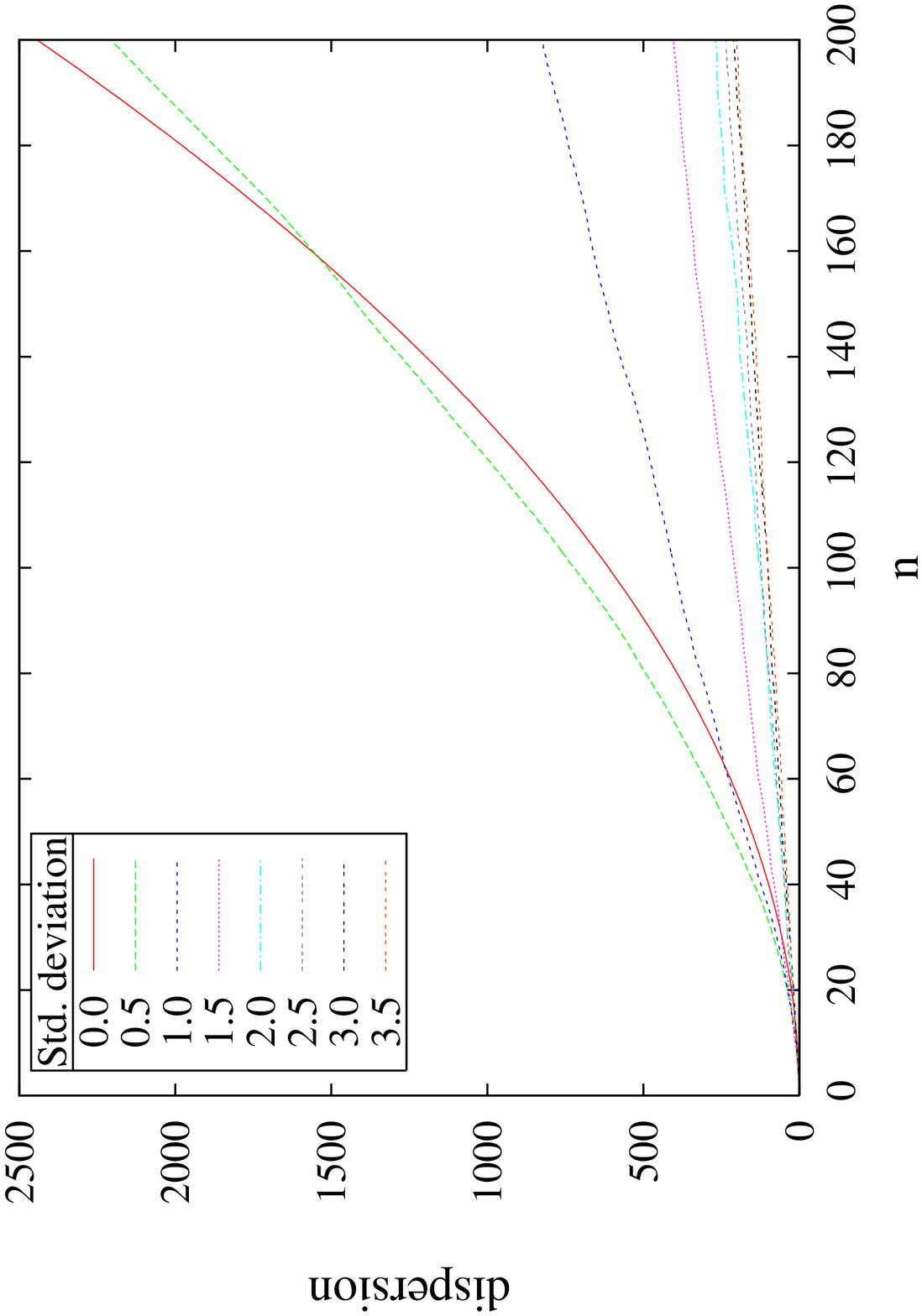} {(Color
online)
  Dispersion $\mathcal{D}_0(D,n)$
  of $n$ steps of QW-RPS in dimension 2 with random phases normally distributed
  around 0 with respective standard deviation. We see a ``continuous'' transition
  between the QW and CRW as function of the standard deviation of the random phase distribution.}

\section{Conclusion}

We have shown that by shifting the amplitudes of the coin register in a quantum
walk by random phases, we can obtain the classical behavior of the quantum walk.
For a Grover coin, the mean probability distribution of such a walk is
equivalent
to the CRW with memory and a biased coin; for the Fourier coin, the mean
probability
distribution is equivalent to the memoryless CRW with an unbiased coin (given
an unsymmetric initial coin state).

The results underlying  Fig.~5 also show how the transition from
QW-RPS to CRW occurs when we increase the dispersion of the normal
distribution of random phases (for the Grover coin). Our results are in a way
complementary to a standard quantization
procedure in physics. Specifically, classical dynamics of physical systems can be canonically quantized, so it is clear
what is the quantum version of a classical process. On the other hand, quantum walk is not obtained by a canonical quantization
procedure from a classical random walk. It is simply defined by a set of instructions that govern the evolution of the quantum
walk. Therefore it is of importance to know what is the underlying classical
process. This underlying process can be reconstructed either by measuring the
coin at each step (cf. \cite{AAKV01a}), or when the quantum walk is subject to
random phase shifts that totally suppress
quantum interference between different evolution paths. 
As a result of the suppression of the quantum interference, the classical
random walk that corresponds to the underlying quantum walk emerges.

\acknowledgements
This research was supported in part by the European Union  projects QAP,
CONQUEST and by the INTAS project 04-77-7289.
In addition this work has been supported   by the Slovak Academy of Sciences via the project CE-PI
I/2/2005 and by the project APVT-99-012304. VB thanks the Alexander von Humboldt Foundation for
support.

\appendix
\section{Dispersion of QW-RPS with Grover coin}
\label{sec:dispersion-qw-rps}
Starting with Eq. (\ref{eq:39}) we can evaluate the dispersion
$\mathcal{D}(D,n)$ of QW-RPS with generalized Grover coin
$G_{r,t}$. The dispersion reads
\begin{equation}
  \label{eq:50}
  \mathcal{D}(D,n)=K\sum_{\mathbf{a}\in D^n}|\sumvec{a_1}{a_n}|^2|\;\Xi(\mathbf{a})|^2
\end{equation}
with $K=\left|\frac{r-t}{\sqrt{|D|}}+\sqrt{|D|}t\right|^2$. Turning
Eq. (\ref{eq:50}) into the recursive relation we obtain
\begin{equation}
  \label{eq:51}
  \begin{split}
    \mathcal{D}(D,n)&=K\sum_{\mathbf{a}\in
      D^{n+1}}\Big\{|\sumvec{a_1}{a_n}|^2\\
    &+2(\sumvec{a_1}{a_n})\cdot\mathbf{e}_{a_{n+1}}+1\Big\}\\
    &\times|\Xi(a_1,\dots,a_n)|^2\;|\Xi(a_n,a_{n+1})|^2\; .
  \end{split}
\end{equation}
We find that $\sum_{a_{n+1}\in D}|\Xi(a_n,a_{n+1})|^2=|r|^2+(|D|-1)|t|^2=1$
for all $a_n\in D$. Hence the first and the last terms in the braces
contribute to Eq. (\ref{eq:51}) with $\mathcal{D}(D,n)+1$. The
middle term has the form
\begin{equation}
  \label{eq:52}
    2K\sum_{\mathbf{a}\in
      D^n}|\Xi(\mathbf{a})|^2\,(\sumvec{a_1}{a_n}) \cdot\sum_{a_{n+1}\in
      D} |\Xi(a_n,a_{n+1})|^2 \mathbf{e}_{a_{n+1}} \; .
\end{equation}
In the sum over $a_{n+1}$ in Eq. (\ref{eq:52}), we can keep just
the terms $a_{n+1}$ such that $\mathbf{e}_{a_{n+1}}$ is parallel with
$\sumvec{a_1}{a_n}$. The  remaining $a_{n+1}$-s cancel out, since for
each such $a_{n+1}$ there is $a_{n+1}'$ such that
$\mathbf{e}_{a_{n+1}}+\mathbf{e}_{a_{n+1}}'=0$. Hence the second term
of Eq. (\ref{eq:52}) can be rewritten as
\begin{equation}
  \label{eq:53}
  \begin{split}
    2K&(|r|^2-|t|^2)\sum_{\mathbf{a}\in
      D^n}|\Xi(\mathbf{a})|^2\,(\sumvec{a_1}{a_n})\cdot\mathbf{e}_{a_n}\\
    &\equiv 2K(|r|^2-|t|^2)R_n\; .
  \end{split}
\end{equation}
The expression for $\mathcal{D}(D,n)$ reads  as
\begin{equation}
  \label{eq:57}
  \mathcal{D}(D,n+1)=\mathcal{D}(D,n)+1+2K(|r|^2-|t|^2)R_n
\end{equation}
with
\begin{equation}
  \label{eq:58}
  \begin{split}
    \mathcal{D}(D,2)&=K\sum_{a_1,a_2\in D}|\Xi(a_1,a_2)|^2|a_1+a_2|^2 \\
    &=K\big\{4|D|\,|r|^2+2(|D|^2-|D|-1)|t|^2\big\}\; .
  \end{split}
\end{equation}
The expression $R_n$ can be rewritten into recursive equation:
\begin{equation}
  \label{eq:54}
  R_{n+1}=(|r|^2-|t|^2)R_n+\frac{1}{K}
\end{equation}
with the initial condition
\begin{equation}
  \label{eq:55}
  \begin{split}
    R_2&=\sum_{a_1,a_2\in
    D}|\Xi(a_1,a_2)|^2(\mathbf{e}_{a_1}+\mathbf{e}_{a_2})\cdot\mathbf{e}_{a_2} \\
  &=2|D||r|^2+(|D|^2-|D|-1)|t|^2\; .
  \end{split}
\end{equation}
Eqs. (\ref{eq:54}) and (\ref{eq:55}) can be solved to obtain
\begin{equation}
  \label{eq:56}
  \begin{split}
    R_n&=\frac{(|r|^2-|t|^2)^{n-2}(KR_2|r|^2-KR_2|t|^2-KR_2+1)-1}{K(|r|^2-|t|^2-1)} \; .
  \end{split}
\end{equation}
Solving Eq. (\ref{eq:57}) and collecting the terms from
Eqs. (\ref{eq:58}),~(\ref{eq:55}), and (\ref{eq:56}) we obtain
\begin{equation}
  \label{eq:59}
  \begin{split}
    \mathcal{D}&(D,n)=\frac{1}{(|r|^2-|t|^2)(1-|r|^2+|t|^2)^2}\times\\
    &\times\Big\{(n-2)\xi-2(|r|^4+|t|^4)+4|r|^2\,|t|^2+\\
    &+(2+\eta)(|r|^2-|t|^2)^n+\eta\big[|t|^2-|r|^2\big]\Big\}\; ,
  \end{split}
\end{equation}
where
\begin{equation}
  \label{eq:60}
  \begin{split}
    \xi&=|r|^2-|r|^6-|t|^2+3|r|^4|t|^2-3|r|^2|t|^4+|t|^6 \; '
  \end{split}
\end{equation}
and
\begin{equation}
  \begin{split}
    \label{eq:61}
    \eta&=2\Big|\frac{r+(|D|-1)t}{\sqrt{|D|}}\Big|(|r|^2-|t|^2-1)\\
    &\times\Big\{2|D||r|^2+(|D|^2-|D|-1)|t|^2\Big\} \; .
  \end{split}
\end{equation}
We may assume that $r$ is real and $t=|t|e^{i\alpha}$. Solving
Eqs. (\ref{eq:34}) and (\ref{eq:35}) we obtain
\begin{eqnarray}
  \label{eq:49}
  |t|&=&\left(\frac{1-r^2}{|D|-1}\right)^{1/2}\; ;\\
  \alpha&=&\pm\arccos\left[\frac{1}{2r}(2-|D|)\left(\frac{1-r^2}{|D|-1}\right)^{1/2}\right]\; ,
\end{eqnarray}
where $|D|\geq4,\,\frac{|D|-2}{|D|}\leq r<1$. Obviously,
$0\leq|r|,|t|\leq1$ and $|r|^2-|t|^2=\frac{|D|r^2-1}{|D|-1}$. Eq. (\ref{eq:59}) contains
only two terms dependent on $n$: $(n-2)\xi$ and $(|r|^2-|t|^2)^n$. The
latter goes to 0 as $n\rightarrow\infty$, hence we get ($n>2$):
\begin{equation}
  \label{eq:62}
    \mathcal{D}(D,n)=\frac{1+|r|^2-|t|^2}{1-|r|^2+|t|^2}(n-2)+O\Big((|r|^2-|t|^2)^n\Big)\; .
\end{equation}

\end{document}